\documentclass{article}

\usepackage{spconf}
\usepackage{amsmath}
\usepackage{graphicx}
\usepackage{amssymb}
\usepackage{xcolor}
\usepackage{colortbl}
\usepackage{booktabs}
\usepackage{algorithm}  
\usepackage{algorithmicx}
\usepackage{algpseudocode}
\usepackage{listings}

\definecolor{darkergreen}{RGB}{21, 152, 56}
\definecolor{red2}{RGB}{252, 54, 65}

\def\onedot{.}
\def\eg{\emph{e.g}\onedot} 
\def\ie{\emph{i.e}\onedot} 
 
\def\etc{\emph{etc}\onedot} 
\def\wrt{\emph{w.r.t}\onedot} 
\def\etal{\emph{et al}\onedot}

\def\eos{\texttt{</s>} }

\title{\textsc{\Large FastEmit:} \uppercase{Low-latency Streaming ASR with \\  Sequence-level Emission Regularization}}
\name{
\parbox{\linewidth}{\centering
Jiahui Yu \; Chung-Cheng Chiu \; Bo Li \; Shuo-yiin Chang \; Tara N. Sainath \; Yanzhang He \\ Arun Narayanan \; Wei Han \;  Anmol Gulati \; Yonghui Wu \; Ruoming Pang}
}
\address{
Google LLC, USA \\
\fontsize{9}{9}\selectfont\ttfamily\upshape
\centering
\{jiahuiyu,rpang\}@google.com
\vspace{-3mm}
}

\begin{document}
\ninept
\maketitle
\begin{abstract}
Streaming automatic speech recognition (ASR) aims to emit each hypothesized word as quickly and accurately as possible. However, emitting fast without degrading quality, as measured by word error rate (WER), is highly challenging. Existing approaches including Early and Late Penalties~\cite{li2020towards} and Constrained Alignments~\cite{sak2015fast, sainath2020emitting} penalize emission delay by manipulating \emph{per-token} or \emph{per-frame} probability prediction in sequence transducer models~\cite{Graves12}. While being successful in reducing delay, these approaches suffer from significant accuracy regression and also require additional word alignment information from an existing model. In this work, we propose a sequence-level emission regularization method, named \emph{FastEmit}, that applies latency regularization directly on \emph{per-sequence} probability in training transducer models, and does not require any alignment. We demonstrate that \emph{FastEmit} is more suitable to the sequence-level optimization of transducer models~\cite{Graves12} for streaming ASR by applying it on various end-to-end streaming ASR networks including RNN-Transducer~\cite{Ryan19}, Transformer-Transducer~\cite{zhang2020transformer, yeh2019transformer}, ConvNet-Transducer~\cite{han2020contextnet} and Conformer-Transducer~\cite{gulati2020conformer}. We achieve \(\mathbf{150\sim300 ms}\) latency reduction with significantly better accuracy over previous techniques on a Voice Search test set. \emph{FastEmit} also improves streaming ASR accuracy from \(4.4\% / 8.9\%\) to \(\mathbf{3.1\% / 7.5\%}\) WER, meanwhile reduces 90th percentile latency from \(210\)ms to only \(\mathbf{30 ms}\) on LibriSpeech.
\end{abstract}

\section{\uppercase{Introduction}}
\label{secs:intro}

End-to-end (E2E) recurrent neural network transducer (RNN-T)~\cite{Graves12} models have gained enormous popularity for streaming ASR applications, as they are naturally streamable~\cite{li2020towards, Ryan19, zhang2020transformer, yeh2019transformer, sainath2020streaming, shuoyiin2020low, yu2020universal, wang2020low}. However, naive training with a sequence transduction objective~\cite{Graves12} to maximize the log-probability of target sequence is \textbf{unregularized} and these streaming models learn to predict better by using more context, causing significant emission delay (\ie, the delay between the user speaking and the text appearing). Recently there are some approaches trying to regularize or penalize the emission delay. For example, Li \etal ~\cite{li2020towards} proposed Early and Late Penalties to enforce the prediction of \eos (end of sentence) within a reasonable time window given by a voice activity detector (VAD). Constrained Alignments~\cite{sak2015fast, sainath2020emitting} were also proposed by extending the penalty terms to each word, based on speech-text alignment information~\cite{variani2018sampled} generated from an existing speech model.

While being successful in terms of reducing latency of streaming RNN-T models, these two regularization approaches suffer from accuracy regression~\cite{li2020towards, sainath2020emitting}. One important reason is because both regularization techniques penalize the \emph{per-token} or \emph{per-frame} prediction probability independently, which is inconsistent with the sequence-level transducer optimization of \emph{per-sequence} probability calculated by the transducer forward-backward algorithm~\cite{Graves12}. Although some remedies like second-pass Listen, Attend and Spell (LAS)~\cite{Chan15} rescorer~\cite{sainath2019two,li2020parallel} and minimum word error rate (MWER) training technique~\cite{rohit2018mwer} have been used to reduce the accuracy regression, these approaches come at a non-negligible compute cost in both training and serving.

In this work, we propose a novel sequence-level emission regularization method for streaming models based on transducers, which we call \emph{FastEmit}. \emph{FastEmit} is designed to be directly applied on the transducer forward-backward per-sequence probability, rather than individual per-token or per-frame prediction of probability independently. In breif, in RNN-T~\cite{Graves12} it first extends the output vocabulary space \(\mathcal{Y}\) with a `blank token' \(\varnothing\), meaning `output nothing'. Then the transducer forward-backward algorithm calculates the probability of each lattice (speech-text alignment) in the \(T \times U\) matrix, where \(T\) and \(U\) is the length of input and output sequence respectively. Finally the optimal lattice in this matrix can be automatically learned by maximizing log-probability of the target sequence. It is noteworthy that in this transducer optimization, emitting a vocabulary token \(y \in \mathcal{Y}\) and the blank token \(\varnothing\) are \emph{treated equally}, as long as the log-probability of the target sequence can be maximized. However, in streaming ASR systems the blank token \(\varnothing\) `output nothing' should be discouraged as it leads to higher emission latency. We will show in detail that \emph{FastEmit}, as a sequence-level regularization method, encourages emitting vocabulary tokens \(y \in \mathcal{Y}\) and suppresses blank tokens \(\varnothing\) across the entire sequence based on transducer forward-backward probabilities, leading to significantly lower emission latency while retaining recognition accuracy.

\emph{FastEmit} has many advantages over other regularization methods to reduce emission latency in end-to-end streaming ASR models: (1) \emph{FastEmit} is a sequence-level regularization based on transducer forward-backward probabilities, thus is more suitable when applied jointly with the sequence-level transducer objective. (2) \emph{FastEmit} does not require any speech-word alignment information~\cite{sainath2020emitting} either by labeling or generated from an existing speech model. Thus it is easy to `plug and play' in any transducer model on any dataset without any extra effort. (3) \emph{FastEmit} has minimal hyper-parameters to tune. It only introduces one hyper-parameter \(\lambda\) to balance the transducer loss and regularization loss. (4) There is no additional training or serving cost to apply \emph{FastEmit}.

We apply \emph{FastEmit} on various end-to-end streaming ASR networks including RNN-Transducer~\cite{Ryan19}, Transformer-Transducer~\cite{zhang2020transformer, yeh2019transformer}, ConvNet-Transducer~\cite{han2020contextnet} and Conformer-Transducer~\cite{gulati2020conformer}. We achieve \(\mathbf{150\sim300 ms}\) latency reduction with significantly better accuracy over previous methods~\cite{sak2015fast, sainath2020emitting, sainath2020streaming} on a Voice Search test set. \emph{FastEmit} also improves streaming ASR accuracy from \(4.4\% / 8.9\%\) to \(\mathbf{3.1\% / 7.5\%}\) WER, meanwhile reduces 90th percentile latency from \(210\)ms to only \(\mathbf{30 ms}\) on LibriSpeech.

\section{\uppercase{Transducer with} \textsc{\large FastEmit}}
\label{secs:transducer_fast_emit}

In this section, we first delve into transducer~\cite{Graves12} and show why naively optimizing the transducer objective is \textbf{unregularized} thus unsuitable for low-latency streaming ASR models. We then propose \emph{FastEmit} as a sequence-level emission regularization method to regularize the emission latency.

\subsection{Transducer}
\begin{figure}[t]
    \centering
    \includegraphics[width=0.9\columnwidth]{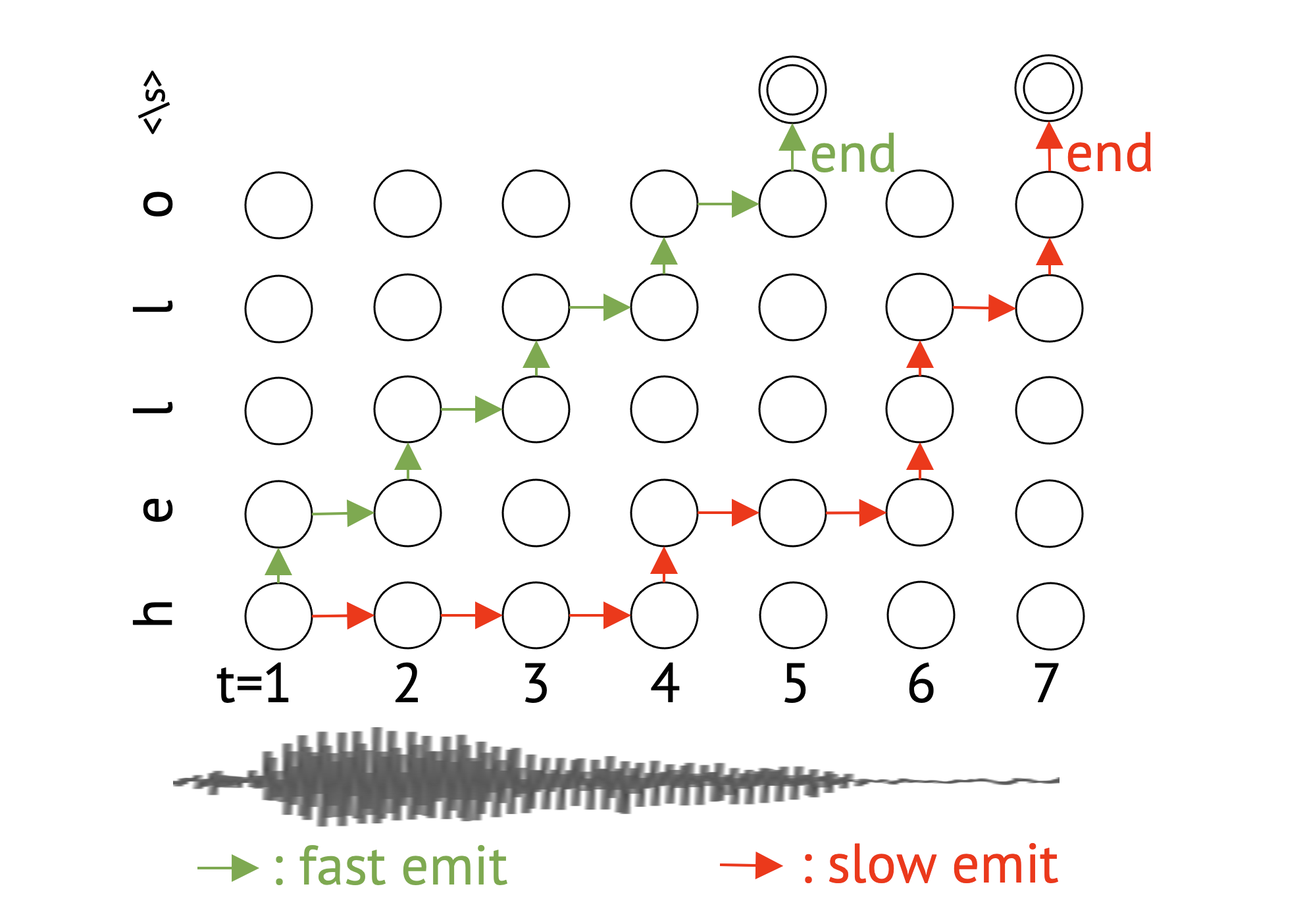}
    \caption{Examples of \textcolor{darkergreen}{fast} and \textcolor{red2}{slow} transducer emission lattices (speech-text alignments). Transducer aims to maximize the log-probability of any lattice, regardless of its emission latency.}
    \label{figs:transducer}
\end{figure}

Transducer optimization~\cite{Graves12} automatically learns probabilistic alignments between an \emph{input sequence} \(\boldsymbol{x} = (x_1, x_2,\ldots, x_T)\) and an \emph{output sequence} \(\boldsymbol{y} = (y_1, y_2,\ldots, y_U)\), where $T$ and $U$ denote the length of input and output sequences respectively. To learn the probabilistic alignments, it first extends the output space \(\mathcal{Y}\) with a `blank token' \(\varnothing\) (meaning `output nothing', visually denoted as right arrows in Figure~\ref{figs:transducer} and ~\ref{figs:fastemit}): \(\mathcal{\bar{Y}} = \mathcal{Y} \cup \varnothing\). The allocation of these blank tokens then determines an alignment between the input and output sequences. Given an input sequence \(\boldsymbol{x}\), the transducer aims to maximize the log-probability of a conditional distribution:
\begin{equation}
\label{eqs:transducer}
\mathcal{L} = -\log P(\hat{\boldsymbol{y}}|\boldsymbol{x}) = -\log \sum_{\boldsymbol{a} \in \mathcal{B}^{-1}(\hat{\boldsymbol{y}})}{P(\boldsymbol{a}|\boldsymbol{x})}
\end{equation}
where \(\mathcal{B}: \mathcal{\bar{Y}} \rightarrow \mathcal{Y}\) is a function that removes the \(\varnothing\) tokens from each alignment lattice \(\boldsymbol{a}\), and \(\hat{\boldsymbol{y}}\) is the ground truth output sequence tokenized from text label.

As shown in Figure~\ref{figs:transducer}, we denote each \emph{node} \((t, u)\) as the probability of emitting the first \(u\) elements of the output sequence by the first \(t\) frames of the input sequence. We further denote the prediction from a neural network \(\hat{y}(t, u)\) and \(b(t, u)\) as the probability of label token (up arrows in figures) and blank token (right arrows in figures) at \emph{node} \((t, u)\). To optimize the transducer objective, an efficient forward-backward algorithm~\cite{Graves12} is used to calculate the probability of each alignment and aggregate all possible alignments before propagating gradients back to \(\hat{y}(t, u)\) and \(b(t, u)\). It is achieved by defining \emph{forward variable} \(\alpha(t, u)\) as the probability of emitting \(\hat{y}[1\mathord{:}u]\) during \(x[1\mathord{:}t]\), and \emph{backward variable} \(\beta(t, u)\) as the probability of emitting \(\hat{y}[u+1\mathord{:}U]\) during \(x[t\mathord{:}T]\), using an efficient forward-backward propagation algorithm:
\begin{align}
\label{eqs:forward}
&\alpha(t, u) = \hat{y}(t, u\mathord{-}1)\alpha(t, u\mathord{-}1) + b(t\mathord{-}1, u)\alpha(t\mathord{-}1, u),\\
\label{eqs:backward}
&\beta(t, u) = \hat{y}(t, u)\beta(t, u\mathord{+}1) + b(t, u)\beta(t\mathord{+}1, u),
\end{align}
where the initial conditions are \(\alpha(1, 0) = 1\), \(\beta(T, U) = b(T, U)\). It is noteworthy that \(\alpha(t, u)\beta(t, u)\) defines the probability of all complete alignments in \(\mathcal{A}_{t,u}: \{\textrm{complete alignment through node} (t, u)\}\):
\begin{align}
\label{eqs:alignment}
P(\mathcal{A}_{t,u}|\boldsymbol{x}) = \sum_{\boldsymbol{a} \in \mathcal{A}_{t,u}}P(\boldsymbol{a}|\boldsymbol{x}) = \alpha(t, u)\beta(t, u).
\end{align}

By diffusion analysis of the probability of all alignments, we know that \(P(\hat{\boldsymbol{y}}|\boldsymbol{x})\) is equal to the sum of \(P(\mathcal{A}_{t,u}|\boldsymbol{x})\) over any top-left to bottom-right diagonal nodes (\ie, all complete alignments will pass through any diagonal cut in the \(T \times U\) matrix in Figure~\ref{figs:transducer})~\cite{Graves12}:
\begin{align}
\label{eqs:loss_alpha_beta}
P(\hat{\boldsymbol{y}}|\boldsymbol{x}) = \sum_{(t,u)\mathord{:}t+u=n} P(\mathcal{A}_{t,u}|\boldsymbol{x}), \forall n : 1 \leq n \leq U + T.
\end{align}
Finally, gradients of transducer loss function \(\mathcal{L} = -\log P(\hat{\boldsymbol{y}}|\boldsymbol{x})\) \wrt~neural network prediction of probability \(\hat{y}(t, u)\) and \(b(t, u)\) can be calculated according to Equations~\ref{eqs:transducer},~\ref{eqs:forward},~\ref{eqs:backward},~\ref{eqs:alignment} and~\ref{eqs:loss_alpha_beta}.

\subsection{\emph{FastEmit}}
\begin{figure}[t]
    \centering
    \includegraphics[width=0.9\columnwidth]{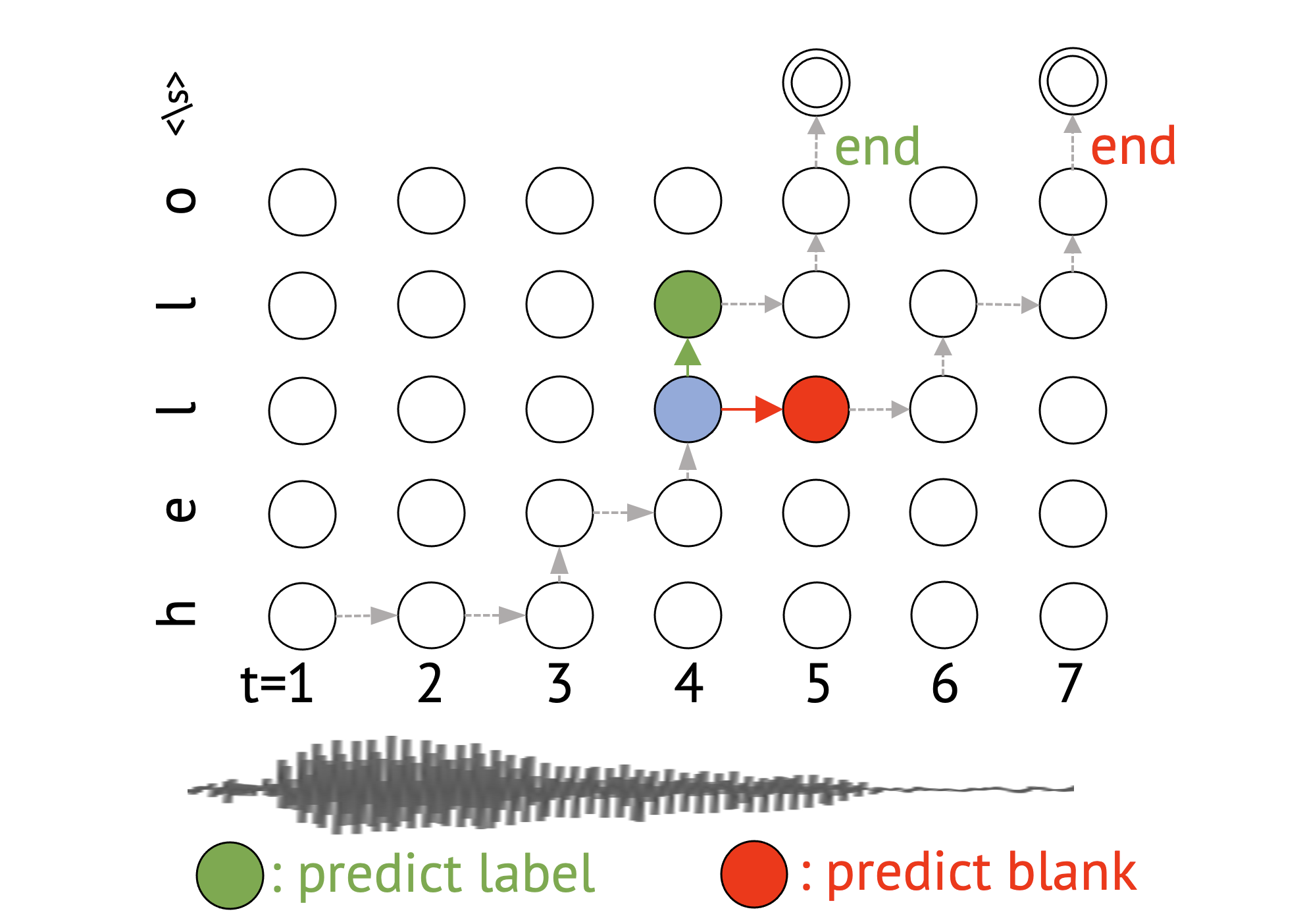}
    \caption{Illustration of \emph{FastEmit} regularization. Consider any node (\eg, \textbf{\textcolor{blue}{blue}} node), \emph{FastEmit} encourages predicting label \(y \in \mathcal{Y}\) (\textbf{\textcolor{darkergreen}{green}} node) instead of predicting blank \(\varnothing\) (\textbf{\textcolor{red2}{red}} node).}
    \label{figs:fastemit}
\end{figure}

Now let us consider any node in the \(T \times U\) matrix, for example, the blue node at \((t, u)\), as shown in Figure~\ref{figs:fastemit}. First we know that the probability of emitting \(\hat{y}[1\mathord{:}u]\) during \(x[1\mathord{:}t]\) is \(\alpha(t, u)\). At the next step, the alignment can either `go up' by predicting label \(u\mathord{+}1\) to the green node with probability \(\hat{y}(t, u)\), or `turn right' by predicting blank \(\varnothing\) to the red node with probability \(b(t, u)\). Finally together with backward probability \(\beta\) of the new node, the probability of all complete alignments \(\mathcal{A}_{t,u}\) passing through node \((t, u)\) in Equation~\ref{eqs:alignment} can be \textbf{decomposed} into two parts:
\begin{align}
\label{eqs:decomposed}
& P(\mathcal{A}_{t,u}|\boldsymbol{x}) = \alpha(t, u)\beta(t, u) = \\
& \underbrace{\alpha(t, u) b(t, u) \beta(t\mathord{+}1, u)}_\text{predict blank} + \underbrace{\alpha(t, u) \hat{y}(t, u) \beta(t, u\mathord{+}1)}_\text{predict label}\nonumber,
\end{align}
which is equivalent as replacing \(\beta(t, u)\) in Equation~\ref{eqs:alignment} with  Equation~\ref{eqs:backward}. From Equation~\ref{eqs:decomposed} we know that gradients of transducer loss \(\mathcal{L}\) \wrt~the probability prediction of any node \((t, u)\) have following properties (closed-form gradients can be found in~\cite{Graves12} Equation 20): 
\begin{align}
\label{eqs:transducer_gradient}
\frac{\partial \mathcal{L}}{\partial{\hat{y}(t, u)}} \propto &~ \alpha(t, u)  \beta(t, u\mathord{+}1)\\
\frac{\partial \mathcal{L}}{\partial{b(t, u)}} \propto &~ \alpha(t, u) \beta(t\mathord{+}1, u).
\end{align}

However, this transducer loss \(\mathcal{L}\) aims to maximize log-probability of all possible alignments, regardless of their emission latency. In other words, as shown in Figure~\ref{figs:fastemit}, emitting a vocabulary token \(y \in \mathcal{Y}\) and the blank token \(\varnothing\) are \emph{treated equally}, as long as the log-probability is maximized. It inevitably leads to emission delay because streaming ASR models learn to predict better by using more future context, causing significant emission delay.

By the decomposition in Equation~\ref{eqs:decomposed}, we propose a simple and effective transducer regularization method, \emph{FastEmit}, which encourages predicting label instead of blank by additionally maximizing the probability of `predict label' based on Equation~\ref{eqs:transducer},~\ref{eqs:loss_alpha_beta} and ~\ref{eqs:decomposed}:
\begin{align}
\label{eqs:fastemit}
&\tilde{P}(\mathcal{A}_{t,u}|\boldsymbol{x}) = \underbrace{\alpha(t, u) \hat{y}(t, u) \beta(t, u\mathord{+}1)}_\text{predict label},\\
&\tilde{\mathcal{L}} = -\log \sum_{(t,u)\mathord{:}t+u=n} (P(\mathcal{A}_{t,u}|\boldsymbol{x}) + \boldsymbol{\lambda} \tilde{P}(\mathcal{A}_{t,u}|\boldsymbol{x})), 
\end{align}
\(\forall n : 1 \leq n \leq U + T\). \(\tilde{\mathcal{L}}\) is the new transducer loss with \emph{FastEmit} regularization and \(\boldsymbol{\lambda}\) is a hyper-parameter to balance the transducer loss and regularization loss. \emph{FastEmit} is easy to implement based on an existing transducer implementation, because the gradients calculation of this new regularized transducer loss \(\tilde{\mathcal{L}}\) follows: 
\begin{align}
\label{eqs:fastemit_gradient}
\frac{\partial \tilde{\mathcal{L}}}{\partial{\hat{y}(t, u)}} &= \mathbf{(1\boldsymbol{+\lambda})} \frac{\partial \mathcal{L}}{\partial{\hat{y}(t, u)}},\\
\frac{\partial \tilde{\mathcal{L}}}{\partial{b(t, u)}} &= \frac{\partial \mathcal{L}}{\partial{b(t, u)}},
\end{align}

To interpret the gradients of \emph{FastEmit}, intuitively it simply means that the gradients of emitting label tokens has a `higher learning rate' back-propagating into the streaming ASR network, while emitting blank token remains the same. We also note that the proposed \emph{FastEmit} regularization method is based on alignment probabilities instead of per-token or per-frame prediction of probability, thus we refer it as \emph{sequence-level emission regularization}.

\section{\uppercase{Experimental Details}}
\label{secs:setup}

\subsection{Latency Metrics}
\begin{figure}[t]
\centering
    \includegraphics[width=0.8\columnwidth]{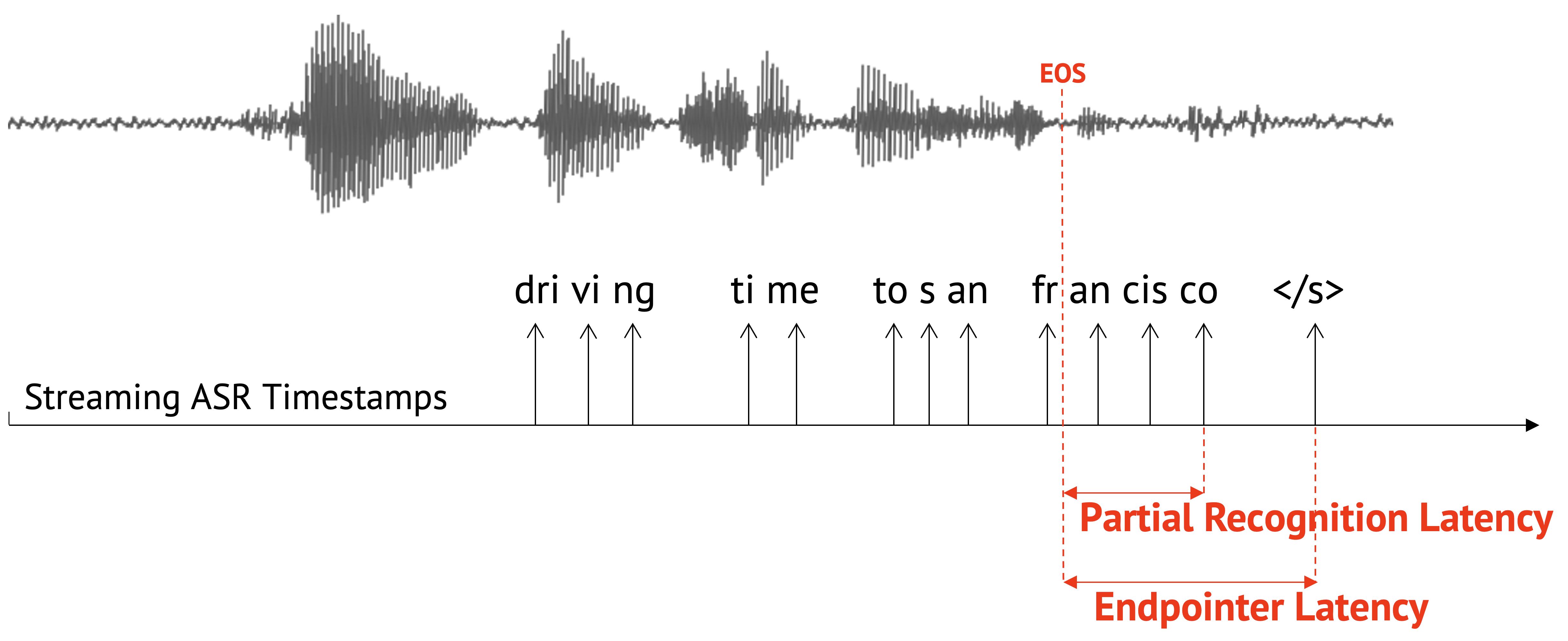}
    \caption{A visual illustration of PR latency and EP latency metrics.}
    \label{figs:latency}
\end{figure}

Our latency metrics of streaming ASR are motivated by real-world applications like Voice Search and Smart Home Assistants. In this work we mainly measure two types of latency metrics described below: (1) partial recognition latency on both LibriSpeech and MultiDomain datasets, and (2) endpointer latency~\cite{chang2017endpoint} on MultiDomain dataset. A visual example of two latency metrics is illustrated in Figure~\ref{figs:latency}. For both metrics, we report both 50-th (medium) and 90-th percentile values of all utterances in the test set to better characterize latency by excluding outlier utterances.

\textbf{Partial Recognition (PR) Latency} is defined as the timestamps difference of two events as illustrated in Figure~\ref{figs:latency}: (1) when the last token is emitted in the finalized recognition result, (2) the end of the speech when a user finishes speaking estimated by forced alignment. PR latency is especially descriptive of user experience in real-world streaming ASR applications like Voice Search and Assistants. Moreover, PR latency is the lower bound for applying other techniques like Prefetching~\cite{shuoyiin2020low}, by which streaming application can send early server requests based on partial/incomplete recognition hypotheses to retrieve relevant information and necessary resources for future actions. Finally, unlike other latency metrics that may depend on hardware, environment or system optimization, PR latency is inherented to streaming ASR models and thus can better characterize the emission latency of streaming ASR. It is also noteworthy that models that capture stronger contexts can emit a hypothesis even before they are spoken, leading to a \textbf{negative PR latency}.

\textbf{Endpointer (EP) Latency} is different from PR latency and it measures the timestamps difference between: (1) when the streaming ASR system predicts the end of the query (EOQ), (2) the end of the speech when a user finishes speaking estimated by forced alignment. As illustrated in Figure~\ref{figs:latency}, EOQ can be implied by jointly predicting the \eos token with end-to-end Endpointing introduced in~\cite{chang2017endpoint}. The endpointer can be used to close the microphone as soon as the user ﬁnishes speaking, but it is also important to avoid cutting off users while they are still speaking. Thus, the prediction of the \eos token has a higher latency compared with PR latency, as shown in Figure~\ref{figs:latency}. Note that PR latency is also a lower bound of EP latency, thus reducing the PR latency is the main focus of this work.

\subsection{Dataset and Training Details}
We report our results on two datasets, a public dataset LibriSpeech~\cite{panayotov2015librispeech} and an internal large-scale dataset MultiDomain~\cite{narayanan2018toward}.

Our main results and ablation studies will be presented on a widely used public dataset LibriSpeech~\cite{panayotov2015librispeech}, which consists of about 1000 hours of English reading speech. For data processing, we extract 80-channel filterbanks feature computed from a 25ms window with a stride of 10ms, use SpecAugment~\cite{park2019specaugment} for data augmentation, and train with the Adam optimizer. We use a single layer LSTM as the decoder. All of these training settings follow the previous work~\cite{han2020contextnet, gulati2020conformer} for fair comparison. We train our LibriSpeech models on 960 hours of LibriSpeech training set with labels tokenized using a 1,024 word-piece model (WPM), and report our test results on LibriSpeech TestClean and TestOther (noisy).

We also report our results a production dataset MultiDomain~\cite{narayanan2018toward}, which consists of 413,000 hours speech, 287 million utterances across multiple domains including Voice Search, YouTube, and Meetings. Multistyle training (MTR)~\cite{li2018multi} is used for noise robustness. These training and testing utterances are anonymized and hand-transcribed, and are representatives of Google’s speech recognition traffic. All models are trained to predict labels tokenized using a 4,096 word-piece model (WPM). We report our results on a test set of 14K Voice Search utterances with duration less than 5.5 seconds long.

\subsection{Model Architectures}
\emph{FastEmit} can be applied to any transducer model on any dataset without any extra effort. To demonstrate the effectiveness of our proposed method, we apply \emph{FastEmit} on a wide range of transducer models including RNN-Transducer~\cite{Ryan19}, Transformer-Transducer~\cite{zhang2020transformer}, ConvNet-Transducer~\cite{han2020contextnet} and Conformer-Transducer~\cite{gulati2020conformer}. We refer the reader to the individual papers for more details of each model architecture. For each of our experiment, we keep the exact same training and testing settings including model size, model regularization (weight decay, variational noise, \etc), optimizer, learning rate schedule, input noise and augmentation, \etc~All models are implemented, trained and benchmarked based on Lingvo toolkit~\cite{shen2019lingvo}.

All these model architectures are based on encoder-decoder transducers. The encoders are based on autoregressive models using uni-directional LSTMs, causal convolution and/or left-context attention layers (no future context is permitted). The decoders are based on prediction network and joint network similar to previous RNN-T models~\cite{li2020towards, Graves12, sainath2020streaming}. For all experiments on LibriSpeech, we report results directly after training with the transducer objective. For all our experiments on MultiDomain, results are reported with minimum word error rate (MWER) finetuning~\cite{rohit2018mwer} for fair comparison.

\section{\uppercase{Results}}
\label{secs:results}
In this section, we first report our results on LibriSpeech dataset and compare with other streaming ASR networks. We next study the  hyper-parameter \(\lambda\) in \emph{FastEmit} to balance transducer loss and regularization loss. Finally, we conduct large-scale experiments on the MultiDomain production dataset and compare \emph{FastEmit} with other methods~\cite{li2020towards, sak2015fast, sainath2020emitting} on a Voice Search test set.

\subsection{Main Results on LibriSpeech}
\begin{table}[ht]
\vspace{-6mm}
\caption{Streaming ASR results on LibriSpeech dataset. We apply \emph{FastEmit} to Large and Medium size streaming ContextNet~\cite{han2020contextnet} and Conformer~\cite{gulati2020conformer}.}
\label{tabs:librispeech}
\centering
\small
\begin{tabular}{@{}l l l l l@{}} \toprule
\textbf{Method} & \textbf{WER} & \textbf{WER} & \textbf{PR50} & \textbf{PR90}\\
 & \textbf{TestClean} & \textbf{TestOther} & \textbf{(ms)} & \textbf{(ms)}\\
\midrule
LSTM & 4.7 & 11.1 & 80 & 180 \\
Transformer & 4.5 & 10.9 & 70 & 190 \\
\midrule
\textbf{Conformer-M} & 4.6 & 9.9 & 140 & 280 \\
\quad +\emph{FastEmit} & 3.7 \textbf{\textsubscript{\textcolor{darkergreen}{(-0.9)}}} & 9.5 \textbf{\textsubscript{\textcolor{darkergreen}{(-0.4)}}} & -40 \textbf{\textsubscript{\textcolor{darkergreen}{(-180)}}} & 80 \textbf{\textsubscript{\textcolor{darkergreen}{(-200)}}}\\
\midrule
\textbf{Conformer-L} & 4.5 & 9.5 & 110 & 230 \\
\quad +\emph{FastEmit} & 3.5 \textbf{\textsubscript{\textcolor{darkergreen}{(-1.0)}}} & 9.1 \textbf{\textsubscript{\textcolor{darkergreen}{(-0.4)}}} & -60 \textbf{\textsubscript{\textcolor{darkergreen}{(-170)}}} & 70 \textbf{\textsubscript{\textcolor{darkergreen}{(-160)}}} \\
\midrule
\textbf{ContextNet-M} & 4.5 & 10.0 & 70 & 270 \\
\quad +\emph{FastEmit} & 3.5 \textbf{\textsubscript{\textcolor{darkergreen}{(-1.0)}}} & 8.6 \textbf{\textsubscript{\textcolor{darkergreen}{(-1.4)}}} & -110 \textbf{\textsubscript{\textcolor{darkergreen}{(-180)}}} & 40 \textbf{\textsubscript{\textcolor{darkergreen}{(-230)}}}\\
\midrule
\textbf{ContextNet-L} & 4.4 & 8.9 & 50 & 210 \\
\quad +\emph{FastEmit} & \textbf{3.1} \textbf{\textsubscript{\textcolor{darkergreen}{(-1.3)}}} & \textbf{7.5}  \textbf{\textsubscript{\textcolor{darkergreen}{(-1.4)}}} & \textbf{-120} \textbf{\textsubscript{\textcolor{darkergreen}{(-170)}}}& \textbf{30} \textbf{\textsubscript{\textcolor{darkergreen}{(-180)}}}\\
\bottomrule
\end{tabular}
\end{table}
We first present results of \emph{FastEmit} on both  Medium and Large size streaming ContextNet~\cite{han2020contextnet} and Conformer~\cite{gulati2020conformer} in Table~\ref{tabs:librispeech}. We did a small hyper-parameter sweep of \(\lambda\) and set \(0.01\) for ContextNet and \(0.004\) for Conformer. \emph{FastEmit} significantly reduces PR latency by \(\mathbf{\sim200 ms}\). It is noteworthy that streaming ASR models that capture stronger contexts can emit the full hypothesis even before they are spoken, leading to a \textbf{negative PR latency}. We also find \emph{FastEmit} even improves the recognition accuracy on LibriSpeech. By error analysis, the deletion errors have been significantly reduced. As LibriSpeech is long-form spoken-domain read speech, \emph{FastEmit} encourages early emission of labels thus helps with vanishing gradients problem in long-form RNN-T~\cite{chiu2019comparison}, leading to less deletion errors.

\subsection{Hyper-parameter \(\lambda\) in \emph{FastEmit}}
\begin{table}[ht]
\vspace{-6mm}
\caption{Study of loss balancing hyper-parameter \(\lambda\) in \emph{FastEmit} on LibriSpeech dataset, based on M-size streaming ContextNet~\cite{han2020contextnet}.}
\label{tabs:librispeech_hp_search}
\centering
\small
\begin{tabular}{@{}l l l l l l@{}} \toprule
\textbf{FastEmit} & \textbf{WER} & \textbf{WER} & \textbf{PR50} & \textbf{PR90}\\
\textbf{H-Param \(\lambda\)} & \textbf{TestClean} & \textbf{TestOther} & \textbf{(ms)} & \textbf{(ms)}\\
\midrule
\textbf{0 \scriptsize{(No FastEmit)}} & 4.5 & 10.0 & 70 & 270  \\
\textbf{0.001} & 4.1 \textbf{\textsubscript{\textcolor{darkergreen}{(-0.4)}}} & 8.7  \textbf{\textsubscript{\textcolor{darkergreen}{(-1.3)}}} & 60 \textbf{\textsubscript{\textcolor{darkergreen}{(-10)}}}& 190 \textbf{\textsubscript{\textcolor{darkergreen}{(-80)}}}\\
\textbf{0.004} & 3.5 \textbf{\textsubscript{\textcolor{darkergreen}{(-1.0)}}} & 8.4  \textbf{\textsubscript{\textcolor{darkergreen}{(-1.6)}}}& -30 \textbf{\textsubscript{\textcolor{darkergreen}{(-100)}}}& 100 \textbf{\textsubscript{\textcolor{darkergreen}{(-170)}}}\\
\textbf{0.008} & 3.6 \textbf{\textsubscript{\textcolor{darkergreen}{(-0.9)}}} & 8.5 \textbf{\textsubscript{\textcolor{darkergreen}{(-1.5)}}} & -80 \textbf{\textsubscript{\textcolor{darkergreen}{(-150)}}}& 50 \textbf{\textsubscript{\textcolor{darkergreen}{(-220)}}}\\
\textbf{0.01} & 3.5 \textbf{\textsubscript{\textcolor{darkergreen}{(-1.0)}}} & 8.6 \textbf{\textsubscript{\textcolor{darkergreen}{(-1.4)}}} & -110 \textbf{\textsubscript{\textcolor{darkergreen}{(-180)}}}& 40 \textbf{\textsubscript{\textcolor{darkergreen}{(-230)}}}\\
\textbf{0.02} & 3.8 \textbf{\textsubscript{\textcolor{darkergreen}{(-0.7)}}} & 9.1 \textbf{\textsubscript{\textcolor{darkergreen}{(-0.9)}}}& -170  \textbf{\textsubscript{\textcolor{darkergreen}{(-240)}}}& -30 \textbf{\textsubscript{\textcolor{darkergreen}{(-300)}}}\\
\textbf{0.04} & 4.4 \textbf{\textsubscript{\textcolor{darkergreen}{(-0.1)}}} & 10.0 \textbf{\textsubscript{\textcolor{darkergreen}{(0.0)}}}& -230  \textbf{\textsubscript{\textcolor{darkergreen}{(-300)}}}& -90 \textbf{\textsubscript{\textcolor{darkergreen}{(-360)}}} \\
\bottomrule
\end{tabular}
\end{table}
Next we study the hyper-parameter \(\lambda\) of \emph{FastEmit} regularization by applying different values on M-size streaming ContextNet~\cite{han2020contextnet}. As shown in Table~\ref{tabs:librispeech_hp_search}, larger \(\lambda\) leads to lower PR latency of streaming models. But when the \(\lambda\) is larger than a certain threshold, the WER starts to degrade due to the regularization being too strong. Moreover, \(\lambda\) also offers flexibility of WER-latency trade-offs.

\subsection{Large-scale Experiments on MultiDomain}
\begin{table}[ht]
\vspace{-6mm}
\caption{Streaming ASR results of \emph{FastEmit} RNN-T, Transformer-T and Conformer-T on a Voice Search test set compared with~\cite{sak2015fast, sainath2020emitting, sainath2020streaming}.}
\label{tabs:voice_search}
\centering
\small
\begin{tabular}{@{}l l l l l l @{}} \toprule
\textbf{Method} & \textbf{WER} & \textbf{EP50} & \textbf{EP90} & \textbf{PR50} & \textbf{PR90}\\
 & & \textbf{(ms)} & \textbf{(ms)} & \textbf{(ms)} & \textbf{(ms)} \\
\midrule
\textbf{RNN-T} & 6.0 & 360 & 750 & 190 & 330 \\
\quad +\emph{CA}~\cite{sak2015fast, sainath2020emitting} & 6.7 \textbf{\textsubscript{\textcolor{red2}{(+0.7)}}} & 450 & 860 & -50 \textbf{\textsubscript{\textcolor{darkergreen}{(-260)}}} & 60 \textbf{\textsubscript{\textcolor{darkergreen}{(-250)}}}\\
\quad +\emph{MaskFrame} & 6.5 \textbf{\textsubscript{\textcolor{red2}{(+0.5)}}} & 250 & 730 & 100 \textbf{\textsubscript{\textcolor{darkergreen}{(-90)}}} & 250 \textbf{\textsubscript{\textcolor{darkergreen}{(-80)}}}\\
\quad +\emph{FastEmit} & 6.2 \textbf{\textsubscript{\textcolor{red2}{(+0.2)}}} & 330 & 650 & -10 \textbf{\textsubscript{\textcolor{darkergreen}{(-200)}}} & 180 \textbf{\textsubscript{\textcolor{darkergreen}{(-150)}}}\\
\midrule
\textbf{Transformer-T} & 6.1 & 400 & 780 & 220 & 370 \\
\quad +\emph{FastEmit} & 6.3 \textbf{\textsubscript{\textcolor{red2}{(+0.2)}}} & 390 & 740 & 60 \textbf{\textsubscript{\textcolor{darkergreen}{(-160)}}} & 220 \textbf{\textsubscript{\textcolor{darkergreen}{(-150)}}}\\
\midrule
\textbf{Conformer-T} & 5.6 & 260 & 590 & 150 & 290 \\
\quad +\emph{FastEmit} & 5.8 \textbf{\textsubscript{\textcolor{red2}{(+0.2)}}} & 290 & 660 & -110 \textbf{\textsubscript{\textcolor{darkergreen}{(-260)}}} & 90 \textbf{\textsubscript{\textcolor{darkergreen}{(-200)}}}\\
\bottomrule
\end{tabular}
\end{table}
Finally we show that \emph{FastEmit} regularization method is also effective on the large scale production dataset MultiDomain. In Table~\ref{tabs:voice_search}, we apply \emph{FastEmit} on RNN-Transducer~\cite{Ryan19}, Transformer-Transducer~\cite{zhang2020transformer} and Conformer-Transducer~\cite{gulati2020conformer}. For RNN-T, we also compare \emph{FastEmit} with other methods~\cite{sak2015fast, sainath2020emitting, sainath2020streaming}. All results are finetuned with minimum word error rate (MWER) training technique~\cite{rohit2018mwer} for fair comparison. In Table~\ref{tabs:voice_search}, \emph{CA} denotes constrained alignment~\cite{sak2015fast, sainath2020emitting}, \emph{MaskFrame} denotes the idea of training RNN-T models with incomplete speech by masking trailing \(n\) frames to encourage a stronger decoder thus can emit faster. We perform a small hyper-parameter search for both baselines \emph{CA} and \emph{MaskFrame} and report their WER, EP and PR latency on a Voice Search test set. \emph{FastEmit} achieves \(\mathbf{150\sim300 ms}\) latency reduction with significantly better accuracy over baseline methods in RNN-T~\cite{Ryan19}, and generalizes further to Transformer-T~\cite{zhang2020transformer} and Conformer-T~\cite{gulati2020conformer}. By error analysis, as Voice Seach is short-query written-domain conversational speech, emitting faster leads to more errors. Nevertheless, among all techniques in Table~\ref{tabs:voice_search}, \emph{FastEmit} achieves best WER-latency trade-off.

\vfill\pagebreak

\bibliographystyle{IEEEbib}
\bibliography{refs}

\begin{thebibliography}{10}

\bibitem{li2020towards}
Bo~Li, Shuo-yiin Chang, Tara~N Sainath, Ruoming Pang, Yanzhang He, Trevor
  Strohman, and Yonghui Wu,
\newblock ``Towards fast and accurate streaming end-to-end asr,''
\newblock in {\em ICASSP 2020-2020 IEEE International Conference on Acoustics,
  Speech and Signal Processing (ICASSP)}. IEEE, 2020, pp. 6069--6073.

\bibitem{sak2015fast}
Ha{\c{s}}im Sak, Andrew Senior, Kanishka Rao, and Fran{\c{c}}oise Beaufays,
\newblock ``Fast and accurate recurrent neural network acoustic models for
  speech recognition,''
\newblock {\em arXiv preprint arXiv:1507.06947}, 2015.

\bibitem{sainath2020emitting}
Tara~N. Sainath, Ruoming Pang, David Rybach, Basi Garc\'{i}a, and Trevor
  Strohman,
\newblock ``{Emitting Word Timings with End-to-End Models},''
\newblock {\em Proc. Interspeech}, 2020.

\bibitem{Graves12}
Alex Graves,
\newblock ``{Sequence Transduction with Recurrent Neural Networks},''
\newblock {\em CoRR}, vol. abs/1211.3711, 2012.

\bibitem{Ryan19}
Yanzhang He, Tara~N. Sainath, Rohit Prabhavalkar, Ian McGraw, Raziel Alvarez,
  Ding Zhao, David Rybach, Anjuli Kannan, Yonghui Wu, Ruoming Pang, Qiao Liang,
  Deepti Bhatia, Yuan Shangguan, Bo~Li, Golan Pundak, Khe~Chai Sim, Tom Bagby,
  Shuo-Yiin Chang, Kanishka Rao, and Alexander Gruenstein,
\newblock ``{Streaming End-to-end Speech Recognition For Mobile Devices},''
\newblock in {\em Proc. ICASSP}, 2019.

\bibitem{zhang2020transformer}
Qian Zhang, Han Lu, Hasim Sak, Anshuman Tripathi, Erik McDermott, Stephen Koo,
  and Shankar Kumar,
\newblock ``Transformer transducer: A streamable speech recognition model with
  transformer encoders and rnn-t loss,''
\newblock in {\em ICASSP 2020-2020 IEEE International Conference on Acoustics,
  Speech and Signal Processing (ICASSP)}. IEEE, 2020, pp. 7829--7833.

\bibitem{yeh2019transformer}
Ching-Feng Yeh, Jay Mahadeokar, Kaustubh Kalgaonkar, Yongqiang Wang, Duc Le,
  Mahaveer Jain, Kjell Schubert, Christian Fuegen, and Michael~L Seltzer,
\newblock ``Transformer-transducer: End-to-end speech recognition with
  self-attention,''
\newblock {\em arXiv preprint arXiv:1910.12977}, 2019.

\bibitem{han2020contextnet}
Wei Han, Zhengdong Zhang, Yu~Zhang, Jiahui Yu, Chung-Cheng Chiu, James Qin,
  Anmol Gulati, Ruoming Pang, and Yonghui Wu,
\newblock ``Contextnet: Improving convolutional neural networks for automatic
  speech recognition with global context,''
\newblock {\em arXiv preprint arXiv:2005.03191}, 2020.

\bibitem{gulati2020conformer}
Anmol Gulati, James Qin, Chung-Cheng Chiu, Niki Parmar, Yu~Zhang, Jiahui Yu,
  Wei Han, Shibo Wang, Zhengdong Zhang, Yonghui Wu, et~al.,
\newblock ``Conformer: Convolution-augmented transformer for speech
  recognition,''
\newblock {\em arXiv preprint arXiv:2005.08100}, 2020.

\bibitem{sainath2020streaming}
Tara~N Sainath, Yanzhang He, Bo~Li, Arun Narayanan, Ruoming Pang, Antoine
  Bruguier, Shuo-yiin Chang, Wei Li, Raziel Alvarez, Zhifeng Chen, et~al.,
\newblock ``A streaming on-device end-to-end model surpassing server-side
  conventional model quality and latency,''
\newblock in {\em ICASSP 2020-2020 IEEE International Conference on Acoustics,
  Speech and Signal Processing (ICASSP)}. IEEE, 2020, pp. 6059--6063.

\bibitem{shuoyiin2020low}
Shuo-Yiin Chang, Bo~Li, David Rybach, Yanzhang He, Wei Li, Tara Sainath, and
  Trevor Strohman,
\newblock ``Low latency speech recognition using end-to-end prefetching,''
\newblock in {\em Interspeech}. ISCA, 2020.

\bibitem{yu2020universal}
Jiahui Yu, Wei Han, Anmol Gulati, Chung-Cheng Chiu, Bo~Li, Tara~N. Sainath,
  Yonghui Wu, and Ruoming Pang,
\newblock ``Universal asr: Unify and improve streaming asr with full-context
  modeling,''
\newblock {\em arXiv preprint arXiv:2010.06030}, 2020.

\bibitem{wang2020low}
Chengyi Wang, Yu~Wu, Shujie Liu, Jinyu Li, Liang Lu, Guoli Ye, and Ming Zhou,
\newblock ``Low latency end-to-end streaming speech recognition with a scout
  network,''
\newblock {\em arXiv preprint arXiv:2003.10369}, 2020.

\bibitem{variani2018sampled}
Ehsan Variani, Tom Bagby, Kamel Lahouel, Erik McDermott, and Michiel Bacchiani,
\newblock ``Sampled connectionist temporal classification,''
\newblock in {\em 2018 IEEE International Conference on Acoustics, Speech and
  Signal Processing (ICASSP)}. IEEE, 2018, pp. 4959--4963.

\bibitem{Chan15}
William Chan, Navdeep Jaitly, Quoc~V. Le, and Oriol Vinyals,
\newblock ``{Listen, Attend and Spell},''
\newblock {\em CoRR}, vol. abs/1508.01211, 2015.

\bibitem{sainath2019two}
Tara~N. Sainath, Ruoming Pang, David Rybach, Yanzhang He, Rohit Prabhavalkar,
  Wei Li, Mirko Visontai, Qiao Liang, Trevor Strohman, Yonghui Wu, Ian McGraw,
  and Chung-Cheng Chiu,
\newblock ``{Two-Pass End-to-End Speech Recognition},''
\newblock {\em Proc. Interspeech}, 2019.

\bibitem{li2020parallel}
Wei Li, James Qin, Chung-Cheng Chiu, Ruoming Pang, and Yanzhang He,
\newblock ``Parallel rescoring with transformer for streaming on-device speech
  recognition,''
\newblock {\em arXiv preprint arXiv:2008.13093}, 2020.

\bibitem{rohit2018mwer}
Rohit Prabhavalkar, Tara~N. Sainath, Yonghui Wu, Patrick Nguyen, Zhifeng Chen,
  Chung-Cheng Chiu, and Anjuli Kannan,
\newblock ``{Minimum Word Error Rate Training for Attention-based
  Sequence-to-Sequence Models},''
\newblock in {\em Proc. ICASSP}, 2018.

\bibitem{chang2017endpoint}
Shuo-Yiin Chang, Bo~Li, Tara~N. Sainath, Gabor Simko, and Carolina Parada,
\newblock ``{Endpoint Detection Using Grid Long Short-Term Memory Networks for
  Streaming Speech Recognition},''
\newblock in {\em Proc. Interspeech}, 2017.

\bibitem{panayotov2015librispeech}
Vassil Panayotov, Guoguo Chen, Daniel Povey, and Sanjeev Khudanpur,
\newblock ``Librispeech: an asr corpus based on public domain audio books,''
\newblock in {\em 2015 IEEE International Conference on Acoustics, Speech and
  Signal Processing (ICASSP)}. IEEE, 2015, pp. 5206--5210.

\bibitem{narayanan2018toward}
Arun Narayanan, Ananya Misra, Khe~Chai Sim, Golan Pundak, Anshuman Tripathi,
  Mohamed Elfeky, Parisa Haghani, Trevor Strohman, and Michiel Bacchiani,
\newblock ``Toward domain-invariant speech recognition via large scale
  training,''
\newblock in {\em Proc. SLT}. IEEE, 2018, pp. 441--447.

\bibitem{park2019specaugment}
Daniel~S Park, William Chan, Yu~Zhang, Chung-Cheng Chiu, Barret Zoph, Ekin~D
  Cubuk, and Quoc~V Le,
\newblock ``Specaugment: A simple data augmentation method for automatic speech
  recognition,''
\newblock {\em arXiv preprint arXiv:1904.08779}, 2019.

\bibitem{li2018multi}
Bo~Li, Tara~N Sainath, Khe~Chai Sim, Michiel Bacchiani, Eugene Weinstein,
  Patrick Nguyen, Zhifeng Chen, Yanghui Wu, and Kanishka Rao,
\newblock ``Multi-dialect speech recognition with a single sequence-to-sequence
  model,''
\newblock in {\em Proc. ICASSP}. IEEE, 2018, pp. 4749--4753.

\bibitem{shen2019lingvo}
Jonathan Shen, Patrick Nguyen, Yonghui Wu, Zhifeng Chen, et~al.,
\newblock ``Lingvo: a modular and scalable framework for sequence-to-sequence
  modeling,''
\newblock {\em arXiv preprint arXiv:1902.08295}, 2019.

\bibitem{chiu2019comparison}
Chung-Cheng Chiu, Wei Han, Yu~Zhang, Ruoming Pang, Sergey Kishchenko, Patrick
  Nguyen, Arun Narayanan, Hank Liao, Shuyuan Zhang, Anjuli Kannan, et~al.,
\newblock ``A comparison of end-to-end models for long-form speech
  recognition,''
\newblock in {\em 2019 IEEE Automatic Speech Recognition and Understanding
  Workshop (ASRU)}. IEEE, 2019, pp. 889--896.

\end{thebibliography}

\end{document}